\pdfoutput=1
\documentclass{llncs}

\usepackage{booktabs}
\usepackage{epsfig,endnotes}
\usepackage{epstopdf}
\usepackage{xcolor}
\usepackage{xspace}
\usepackage{enumitem}
\usepackage{url}
\usepackage{breakurl}
\usepackage{listings}
\usepackage{color}
\usepackage[tableposition=top,font={footnotesize, bf},figurename=Fig., skip=7pt]{caption}     
\usepackage{graphicx}

\newcommand{\point}[1]{\par\smallskip\noindent\textbf{#1} }

\definecolor{dkgreen}{rgb}{0,0.6,0}
\definecolor{gray}{rgb}{0.5,0.5,0.5}

\newcommand{\eg}{e.g., }
\newcommand{\ie}{i.e., }
\newcommand{\etal}{et al. }

\newcommand{\topList}{150K\xspace}
\newcommand{\sitesWithSW}{7,444\xspace}
\newcommand{\sitesWithSWsecond}{9,383\xspace}
\newcommand{\sitesWithSWperc}{4.96\%\xspace}
\newcommand{\sitesWithSWsecondperc}{6.25\%\xspace}

\newcommand{\swNoParty}{336\xspace}
\newcommand{\swNoPartyperc}{4.51\%\xspace}
\newcommand{\swThirdParty}{5,054\xspace}
\newcommand{\swThirdPartyperc}{67.89\%\xspace}
\newcommand{\swFirstParty}{2,054\xspace}
\newcommand{\swFirstPartyperc}{27.59\%\xspace}

\newcommand{\swForAns}{164\xspace}
\newcommand{\swForAnsperc}{3.24\%\xspace}
\newcommand{\swForAds}{3,289\xspace}
\newcommand{\swForAdsperc}{44.18\%\xspace}
\newcommand{\swtpToAdsperc}{65.08\%\xspace}
\newcommand{\swtpNOAdsperc}{34.92\%\xspace}

\newcommand{\adoptionIncrease}{26\%\xspace}
\newcommand{\swslong}{Service Workers\xspace}
\newcommand{\swlong}{Service Worker\xspace}
\newcommand{\sw}{\emph{SW}\xspace}
\newcommand{\sws}{\emph{SWs}\xspace}

\newcommand{\oneTp}{51.6\%\xspace}
\newcommand{\noTp}{32.1\%\xspace}
\newcommand{\twoMoreTp}{16.3\%\xspace}

\author{
George Pantelakis\inst{1}
\and Panagiotis Papadopoulos\inst{2}\\
Nicolas Kourtellis\inst{2}
\and Evangelos P. Markatos\inst{1}}

\institute{FORTH/University of Crete,\\
\and
Telefonica Research, Barcelona, Spain}

\authorrunning{G. Pantelakis et al.}

\begin{document}
\title{Measuring the (Over)use of Service Workers for In-Page Push Advertising Purposes}

\titlerunning{Measuring the (Over)use of Service Workers for In-Page Push Advertising} 

\maketitle
\begin{abstract}
Rich offline experience, periodic background sync, push notification functionality, network requests control, improved performance via requests caching are only a few of the functionalities provided by the \swlong (\sw) API.
This new technology, supported by all major browsers, can significantly improve users' experience by providing the publisher with the technical foundations that would normally require a native application.
Albeit the capabilities of this new technique and its important role in the ecosystem of Progressive Web Apps (PWAs), it is still unclear what is their actual purpose on the web, and how publishers leverage the provided functionality in their web applications.

In this study, we shed light in the real world deployment of \sws, by conducting the first large scale analysis of the prevalence of \sws in the wild.
We see that \sws are becoming more and more popular, with the adoption increased by \adoptionIncrease only within the last 5 months. 
Surprisingly, besides their fruitful capabilities, we see that \sws are being mostly used for In-Page Push Advertising, in \swtpToAdsperc of the \sws that connect with 3rd parties.
We highlight that this is a relatively new way for advertisers to bypass ad-blockers and render ads on the user's displays natively. 

\keywords{Service Workers, Push Ads, Push Notification Advertising}
\end{abstract}

\section{Introduction}
\label{sec:intro}

The proliferation of, and our ever-increasing reliance on, the Web have boosted the development of more complex and user-friendly Web applications that can operate cross-platform (on both desktop and mobile Web).
Recent advancements in the contemporary browsers and in the availability of technologies like the \sw API have 1) enabled users to receive timely updates via push notifications, 2) their content synced on the background, 3) improved performance (via request caching) and 4) even allowed to work offline. 

These rich capabilities of \sws played an important role in the birth and growth of a whole separate type of application software called Progressive Web Apps (PWAs)~\cite{webapps}.
PWAs are built on top of three requirements: HTTPS, \sws and a web app manifest.
By combining functionalities of different web APIs (\eg WebRTC, Cache API, Push API), PWAs are capable of providing the benefits of both native apps and websites worlds: reliability, rich user experience, and multi-platform support via a single codebase~\cite{webapps2}.

The somewhat revolutionary functionality of \sws could not avoid drawing the attention of the academic community with regards to its security aspects.
Specifically, research studies have shown that this technology provides rich capabilities not only to users and web developers, but to potential attackers as well.
In~\cite{papadopoulosmaster}, authors present a framework that exploits \sws functionality to launch attacks like DDoS, cryptojacking and distributed password cracking.
In~\cite{karami2021awakening}, authors investigate the potential privacy leaks that malicious \sws can cause on a victim's browser.

Notwithstanding the important research on the \sw API, yet it is still unknown what is the prevalence and the growth of the \sw deployment across the Web and how publishers leverage the provided functionality of \sws in their Web applications.
In this study, we aim to address these exact questions, by conducting a full-scale analysis of \sws (the core component of PWAs) in the wild. 
Specifically, we crawl a large number of websites to detect the deployment of \sws, monitor and characterize their communications across the Internet, and investigate their purpose of existence and operation on the websites found.

In summary, the contributions of our present work are:
\begin{enumerate}
    \item By crawling the top \topList sites of the Tranco list, we detect a dataset of \sitesWithSW \sws-registering websites. The same crawl after 5 months reveals a high increase (26\%) in the adoption of \sws.
    \item We use Wayback Machine to go back in time and find that, from 2015 till today, there were $1.62\times$ more publishers per year, on average, utilizing \sws in their web applications.
    \item By analysing our collected dataset, we conduct the first full-scale study of the \sws deployment on the Web. Specifically, we investigate with whom the deployed \sws communicate over the Internet, what are the websites that use such technology the most, as well as what is the purpose of the deployed \sws. Surprisingly, we see that despite the important functionality of \sws (\eg timely notifications, background sync, etc.), yet a stunning \swtpToAdsperc of the \sws that connect with 3rd parties use \sws for pushing ads to the users, under the radar of possibly deployed ad-blockers.
\end{enumerate}
\section{\swslong}
\label{sec:background}

A \swlong is a JavaScript script that runs separately from the main browser thread, and can intercept network requests, perform caching or retrieving resources from the cache, and deliver push messages.
\sws are independent from the Web application they are associated with, so they cannot access the DOM directly.
\sws are non-blocking and fully asynchronous.
Therefore, synchronous XHR and localStorage cannot be used inside a \sw. 
Also, a \sw can import and execute 3rd party scripts within its context, and receive push messages from a remote server, thus letting the associated website push notifications to the user (even when the website is not open in a browser tab).
Finally, a \sw can be registered to the browser via the {\tt serviceWorkerContainer.register()} or {\tt navigator.serviceWorker.register()} function, which  take  as  argument  the  (HTTPS only) URL  of  the  remote  JavaScript file that contains the  worker's  script.
This  URL  is  passed  to the internal  browser’s  engine  and  is  fetched  from  there.
For security purposes, this JavaScript file can be fetched only from the first-party domain, \ie cannot be hosted by a CDN or a 3rd party server.

\begin{figure}[t]
    \centering
    \begin{minipage}[t]{0.47\textwidth}
        \centering
        \includegraphics[width=1.06\textwidth]{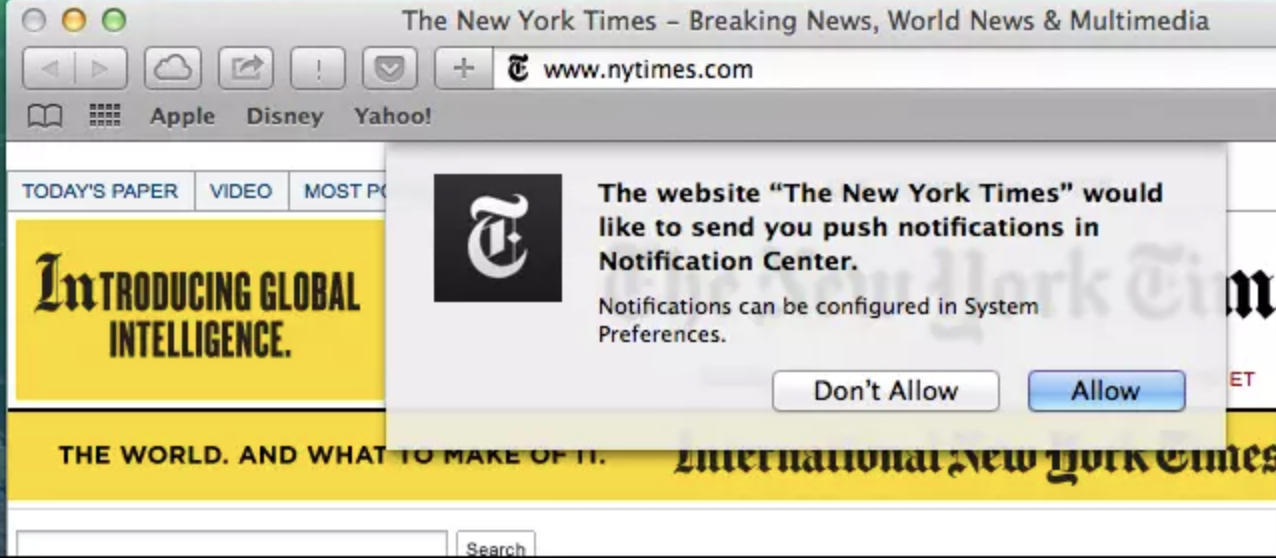}
        \caption{The Web Push API enables developers to deliver asynchronous notifications and updates to users who opted-in.}
        \label{fig:pushApi}
    \end{minipage}
    \hfill
    \begin{minipage}[t]{0.47\textwidth}
        \centering
        \includegraphics[width=1.06\textwidth]{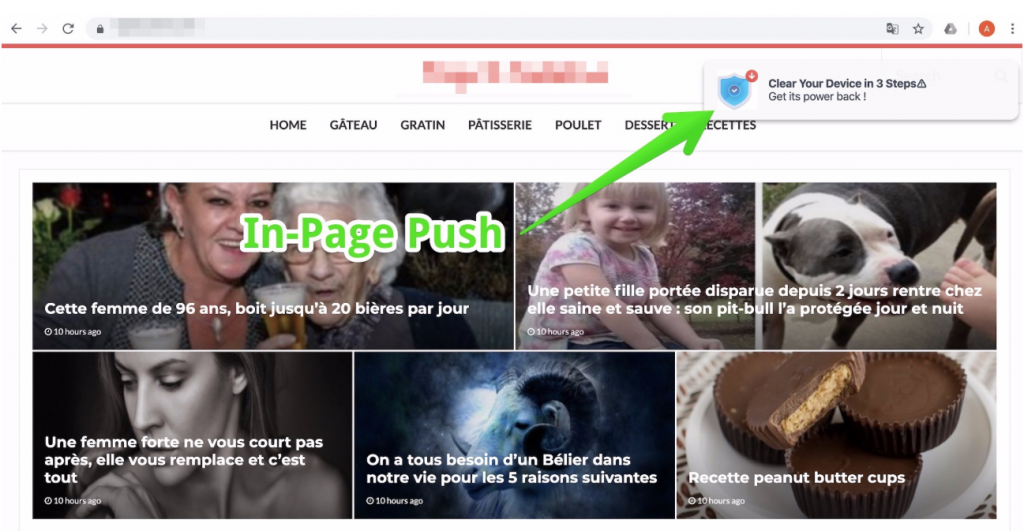}
        \caption{An in-page push advertisement as it appears on the user's screen on top of other windows.}
        \label{fig:pushad}
    \end{minipage}
\end{figure}

\subsection{Web Push Notifications}
The Web Push API gives web applications the ability to receive messages pushed from a remote server, whether or not the Web app is in the foreground, or even loaded in a browser tab.
As shown in Figure~\ref{fig:pushApi}, the Web Push API enables developers to deliver asynchronous notifications and updates to (desktop or mobile) users that opt-in, resulting in better engagement with timely new content.
For an app to receive push messages, it has to have an active \sw and subscribe to push notifications (each subscription is unique to a \sw).
The endpoint for the subscription is a unique \emph{capability URL}, and the knowledge of the endpoint is all that is necessary to send a message to the application's users. 
Therefore, the endpoint URL needs to be kept secret or anyone might be able to send push messages to the app's users.

\subsection{In-Page Push Advertising}
Web push notification technology itself is nothing new, but it has started to be used for advertising purposes very recently.
In fact, push marketing skyrocketed at the end of 2018~\cite{pushads2018}.
Push ads are a type of native ad format in the form of a notification message from a website, which appears on the user's screen on top of other windows as shown in Figure~\ref{fig:pushad}.
Users who click on those messages get redirected to the advertiser's landing page, thus, generating ad-conversion.

The in-page push ad delivery is cross-platform and aims to offer an \emph{opt-in based}, highly engaging way for advertisers to reconnect and expand their audiences, while at the same time it achieves higher click-through and conversion rates than other ad formats~\cite{BravePush}.
A push notification usually consists of: (i) the main image which conveys the sense of the ad impression, (ii) the small icon which explains the main image, (iii) the headline which is the main element to engage users and (iv) the message text that shows the main details of the offer.
Contrary to traditional programmatic advertising~\cite{papadopoulos2018cost,castelluccia2014selling,pachilakis2019no}, in push ads, advertisers pay for clicks (\ie Cost-Per-Click) and not for impressions (\ie Cost-per-Impression).
The minimum cost per click starts from \$0.0104~\cite{pushPrice}, but in Real-Time Bidding the median cost per impression has been measured to be as low as \$0.0025~\cite{papadopoulos2017if}.

\begin{figure}[t]
    \centering
    \includegraphics[width=0.8\textwidth]{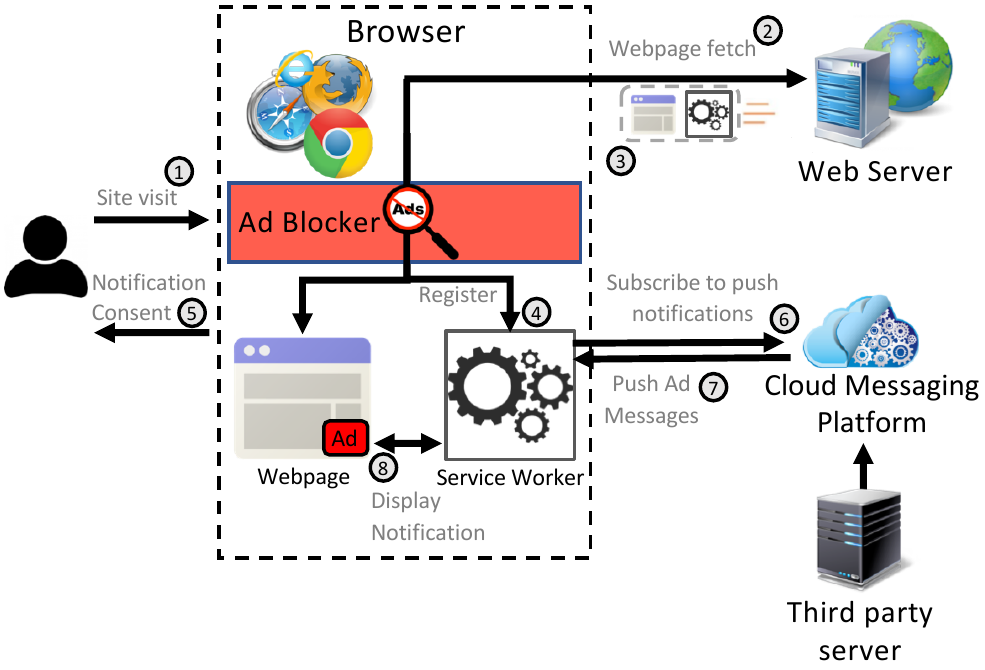}\vspace{-0.1cm}
    \caption{High level overview of how \sws deliver push ads on the user display even with ad blocker deployed.}
    \label{fig:sw}
\end{figure}
\section{Use Case}
\label{sec:usecase}

In Figure~\ref{fig:sw}, we present a high level overview of how \sws and push notifications work.
As we can see, first (step 1), the user visits a website they are interested in, thus, instructing a browser to connect with a web server (step 2) that responds back with the web page's HTML/CSS/JavaScript resources, along with a \sw script, which gets registered (step 3).
This snippet will deploy a \sw inside the user's browser (step 4) which operates independently from the rendered website.
Then, the \sw will ask the user's permission to push notification massages on their display (step 5) and if granted, it will establish a communication channel with a remote messaging platform to subscribe to their push notifications (step 6).
Whenever the message publishing entity (\eg news update feed server, article recommendation server, ad server) behind the messaging platform has updates to push to the website's users, it uploads them to the platform which will push them to all subscribed users (step 7).
On the user's end, upon message arrival, the deployed \sw creates a push notification with the received message on the user's display (step 8).
As shown, a \sw may establish a separate communication channel with a remote messaging platform that cannot be monitored or filtered by any potentially deployed ad-blocking browser extension.
This means that whenever a user opts-in to receive updates from a website, they may start receiving ad notifications instead, even if they have an ad-blocker deployed.

\section{Data Collection}
\label{sec:dataset}

 \begin{table}[t]
    \centering
    \caption{Summary of our dataset}
    \begin{tabular}{lr}
    \toprule
    {\bf Data} & {\bf Volume} \\
    \midrule
        Websites parsed & \topList \\
        \emph{(1st crawl, 12.20)} Websites registering a \sw (SW)  & \sitesWithSW(\sitesWithSWperc) \\ \hline
        SWs that do not communicate with any remote server & \swNoParty(\swNoPartyperc) \\ 
        SWs communicating only with the first party & \swFirstParty(\swFirstPartyperc) \\ 
        SWs communicating with at least one 3rd party & \swThirdParty(\swThirdPartyperc) \\ \hline
        SWs communicating with at least one ad server & \swForAds(\swForAdsperc) \\ 
        SWs communicating with at least one analytics server & \swForAns(\swForAnsperc) \\ \hline
         \emph{(2nd crawl, 05.21)} Websites registering a SW & \sitesWithSWsecond(\sitesWithSWsecondperc) \\
         \bottomrule
    \end{tabular}
    \label{tab:summary}
\end{table}

\point{Crawling infrastructure} After manual inspection, we see that there are websites checking first if the site has push notification permissions, before registering \sws.
This means that in order to perform a large scale crawl of websites and detect the deployment of \sws, and the use of push notifications, some sort of automation for the notification consent is required.
To address this, we leverage the crawler presented in~\cite{10.1145/3419394.3423631}.
This crawler creates docker containers with fresh instrumented Chromium browser instances and browser automation scripts.
The browser has the \emph{RequestPermission} and \emph{PermissionDecided} methods of the class \emph{Permission-ContextBase}, modified to automatically grant permissions on every site.
Then, a custom Puppeteer~\cite{puppeteer} script listening to \emph{serviceworkercreated} event is used to log when a \sw is registered by a website, the page that registered this \sw and the URI of the source code.
As soon as a \sw is registered, it can subscribe for push notifications via a Cloud Messaging Platform (\eg Firebase Cloud Messaging~\cite{firebase}) with an API key passed from the server to the browser, which is also logged by listening for \emph{PushManager.subscribe} events.
Then, the custom Puppeteer script logs the communication between the \sw and the Web.

\point{Creating the Dataset} We create a dataset of websites that utilize \sws, by crawling the landing pages of the \topList top sites of a (deduplicated, pay-level only domains) Tranco list~\cite{customTrancoList} in December 2020.
Each site is visited for three minutes, during which, and according to our experiments, is sufficient for a \sw to be register and make the first touch with the corresponding server(s) (\emph{1st crawl}).
After 5 months (April 2021), we revisit the websites of our initial Tranco list to inspect how the ecosystem evolved, \ie sites dropping \sws, new sites adopting \sws (\emph{2nd crawl}).
Table~\ref{tab:summary} summarizes the data collected.

\point{Data analysis}
To classify the network traffic of the registered \sws in our data, we use the 1Host filterlist~\cite{1Hosts} and flag the ad-related domains in our weblogs. 
Additionally, we used SimilarWeb~\cite{similarweb} to categorize the sites registering \sw based on the content they deliver.

\begin{figure}[t]
    \centering
    \includegraphics[width=0.8\textwidth]{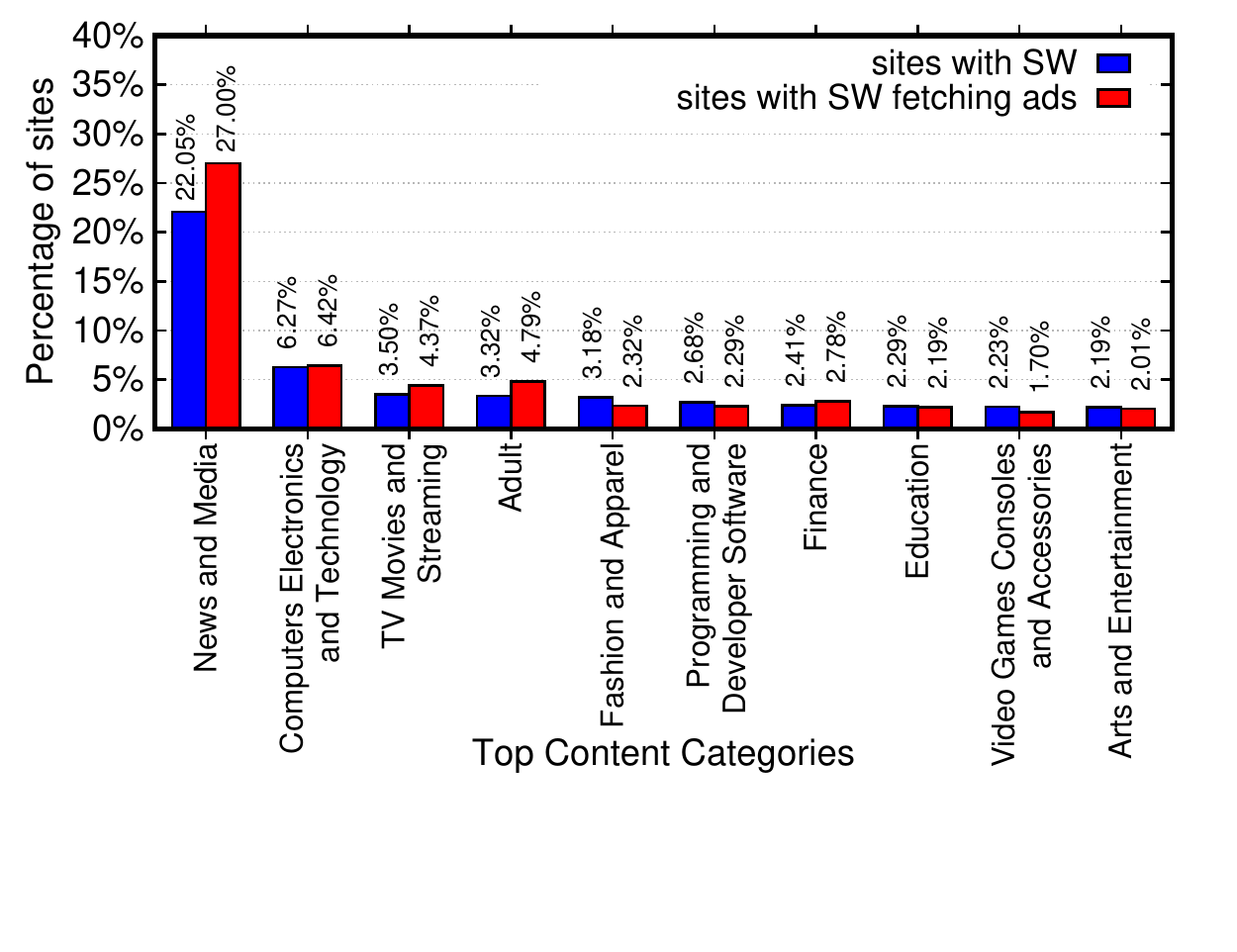}\vspace{-1cm}
    \caption{Top 10 content categories of sites using (i) \sws (blue) and (ii) \sws that communicate with advertisers (red). As we can see, \sws is a technique widely used in sites delivering content related to `News and Media' and `Computers Electronics and Technology'.}\vspace{-.2cm}
    \label{fig:sitesSWcontent}
\end{figure}

\point{Historical analysis and static detection}
To explore the evolution of \sws across time, we use our set of \sw-registering sites to extract heuristics and keywords that indicate the registration or use of \sws.
This way, we develop a crawler that can statically detect the registration of \sws in this set of sites.
Next, we use Wayback Machine~\cite{waybackMachine}, and specifically {\tt waybackpy} Python package~\cite{waybackpy} to go back in time and find the day that these websites started deploying \sws in their visitors' browsers.

\begin{figure}[t]
    \centering
    \includegraphics[width=.65\textwidth]{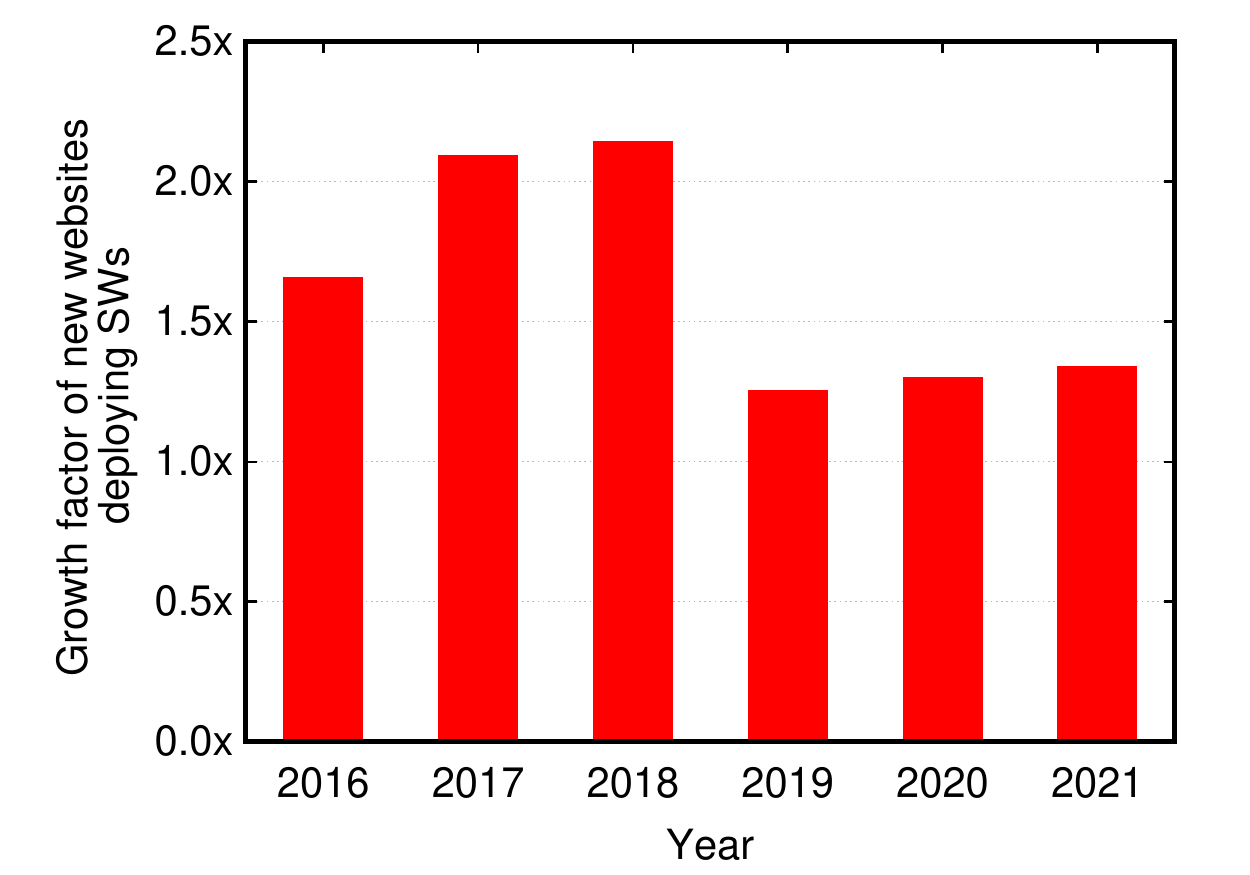}
    \caption{Growth factor of the \sws deployment in our dataset. From 2015 till mid 2021, there are $1.62\times$ more publishers per year, on average, utilizing \sws in their Web apps.}
    \label{fig:archiveMesurments}  
\end{figure}

\begin{figure}[t]
    \centering
    \includegraphics[width=.65\textwidth]{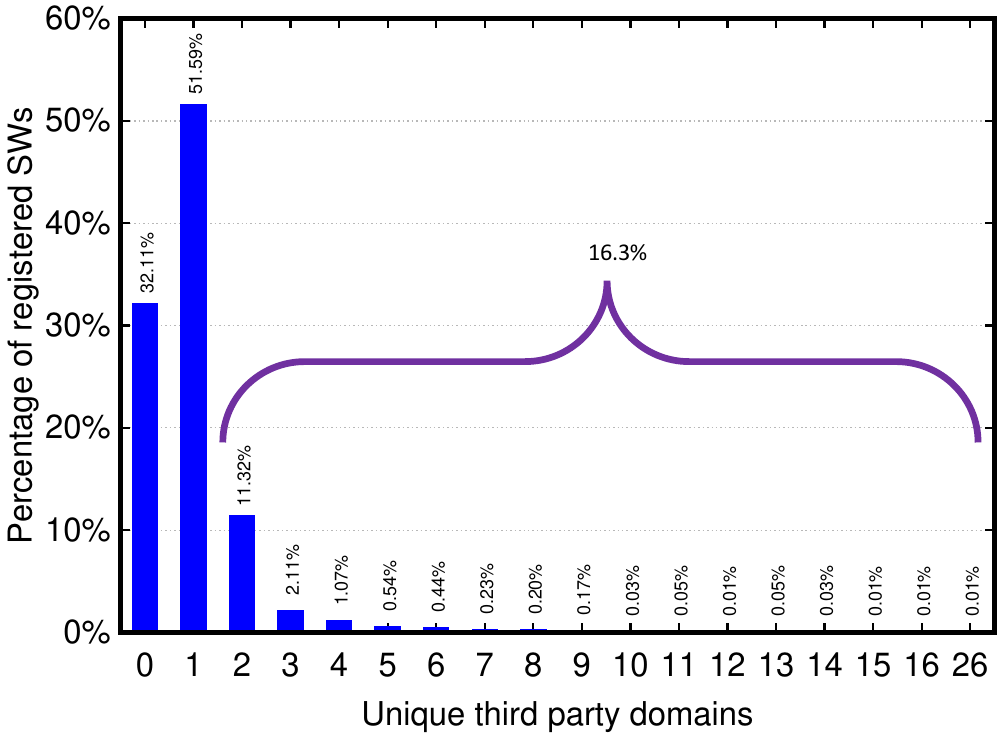}
    \caption{The number of distinct 3rd parties each \sw communicates with. \noTp of the \sws do not communicate with a 3rd party. The majority (\oneTp) communicates with exactly one 3rd party. \twoMoreTp communicate with 2 or more (even 26!) distinct 3rd parties.}
    \label{fig:thirdParties}
\end{figure}

\begin{figure*}[t]
    \centering
    \begin{minipage}[t]{0.49\textwidth}
        \centering
        \includegraphics[width=1.0\textwidth]{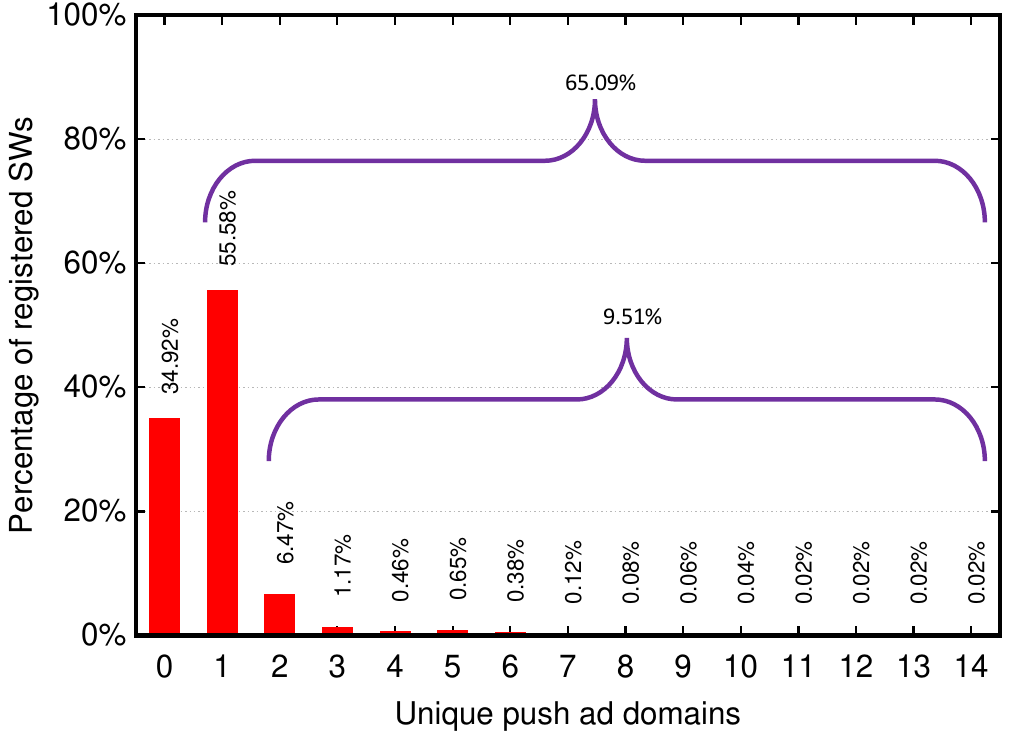}
        \caption{Number of distinct push ad 3rd parties each \sw communicates with. \swtpToAdsperc of \sws communicate with 3rd parties and receive content from at least one push advertiser.
        \swtpNOAdsperc of \sws perform at least one request to 3rd parties but communicate with zero ad servers.}
        \label{fig:NumOfAdServers}
    \end{minipage}
    \hfill
    \begin{minipage}[t]{0.49\textwidth}
        \hspace{-0.2cm}\includegraphics[width=1.06\textwidth]{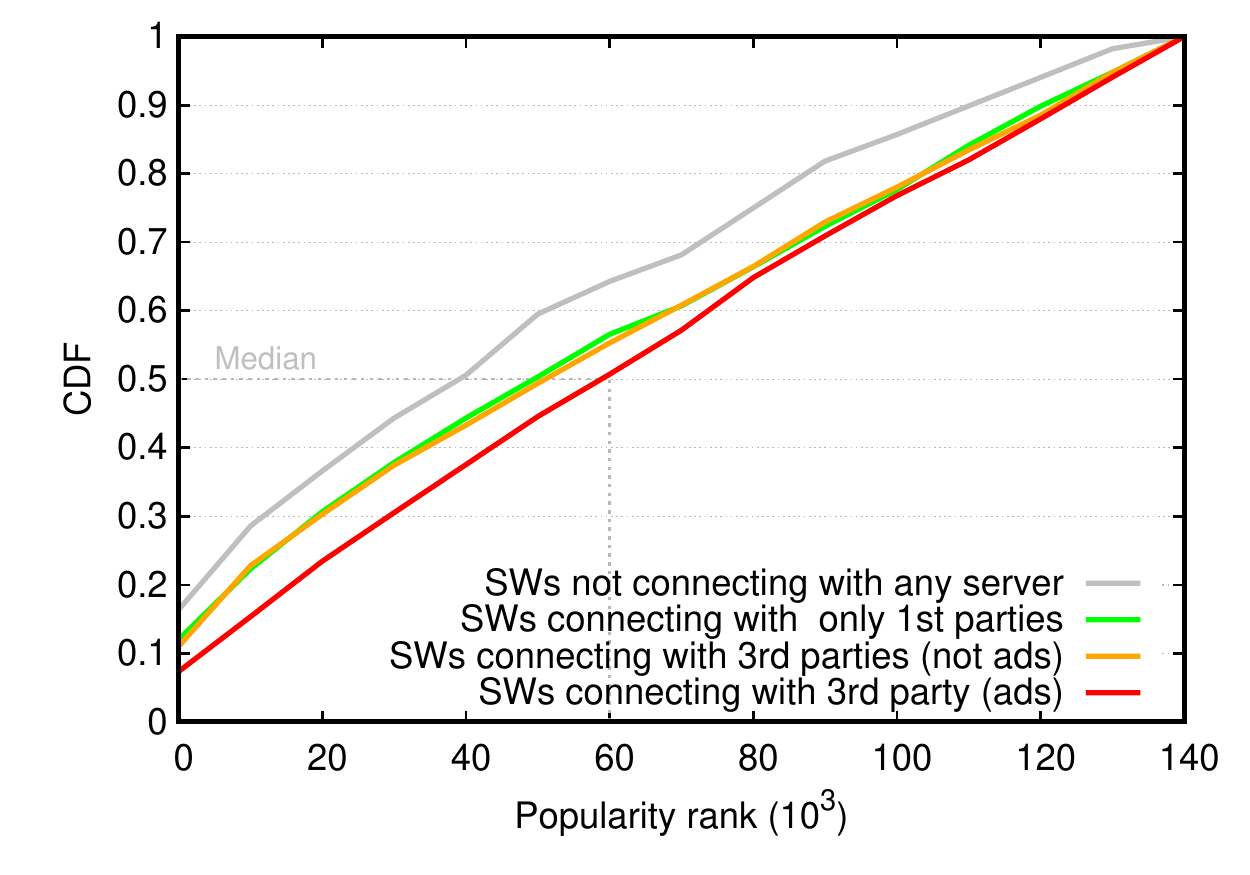}
        \caption{Cumulative distribution function of the popularity rank of the websites with \sws. As we see, the sites that tend to deploy ad pushing \sws are less popular than the ones that use \sw only locally, and do not connect to any remote 3rd party server.}
        \label{fig:rank}
    \end{minipage}
\end{figure*}

\point{Ethical Considerations}
It is important to note that during the conduct of this study, we neither gathered or used any user data, nor impeded or tampered with the proper operation of the sites we crawled, in any way.
Our research was purely limited in passively monitoring the behavior of \sws.

 
\section{Measurements}
\label{sec:analysis}

\point{What are the kind of sites deploying \sws?}
By crawling the top \topList websites of the Tranco list, we find that \sitesWithSW(\sitesWithSWperc) of these sites register one or more \sws in the users' browser (Table~\ref{tab:summary}).
To understand what kind of sites deploy such a technique, we query Similarweb~\cite{similarweb} for the content category of each of our sites and we get a response for 86.7\% of them.
In Figure~\ref{fig:sitesSWcontent}, we plot the top-10 categories.
We see that \emph{the sites that mostly use \sws (in blue) are related to `News and Media' (22.05\%)}, with the categories of `Computers Electronics and Technology' and `Arts and Entertainment' following (6.27\% and 3.5\%, respectively).

\point{What is the prevalence of \sw deployment?}
By revisiting the sites of the same Tranco list after 5 months (as described in Section~\ref{sec:dataset}) via the same crawler, we find a total of \sitesWithSWsecond websites registering a \sw (\sitesWithSWsecondperc of the total sites crawled), which indicates a \adoptionIncrease increase.
More specifically, we find (i) 6,173 websites using \sws in both crawls, (ii) 1,271 websites that stopped using them at some time after our 1st crawl, and (ii) 3,210 new websites deploying them in their visitors' browsers.

This rapid growth in the prevalence of \sws within just 5 months, motivated us to go back in time and observe their evolution across the years.
Specifically, by using Wayback Machine web archive~\cite{waybackMachine} and our initial set of \sw-registering websites, we crawled previous versions of their landing pages to spot when they started using \sws.
As a result, we crawled all the way back to 2015, when the first websites in our dataset started using \sws.
As seen in Figure~\ref{fig:archiveMesurments}, after 2015, every year we observe an average growth factor of 1.62.
In 2017 and 2018 we see this growth increasing, with $2.09\times$ and $2.14\times$ more websites deploying \sws than the previous year, respectively.

\begin{figure}[t]
    \centering
    \begin{minipage}[t]{0.47\textwidth}
        \centering
        \includegraphics[width=1.03\textwidth]{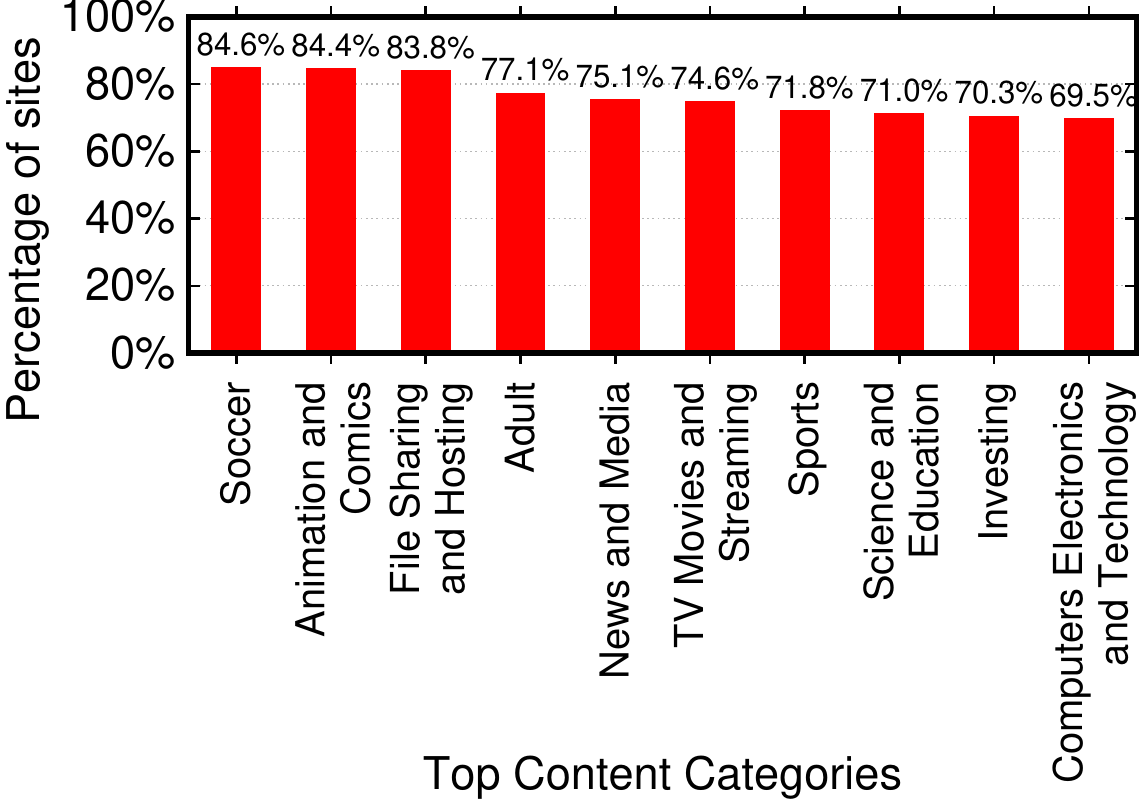}
        \caption{Breakdown the top-10 categories for sites using \sws to serve ads.
        `Soccer' sites are the most aggressive in using \sws for advertising (84.62\%), with `Animation and Comics' and `File Sharing and Hosting' following. 'News and Media' sites are next (75\%).}
        \label{fig:adcontent}
    \end{minipage}
    \hfill
    \begin{minipage}[t]{0.47\textwidth}
        \centering
        \includegraphics[width=1\textwidth]{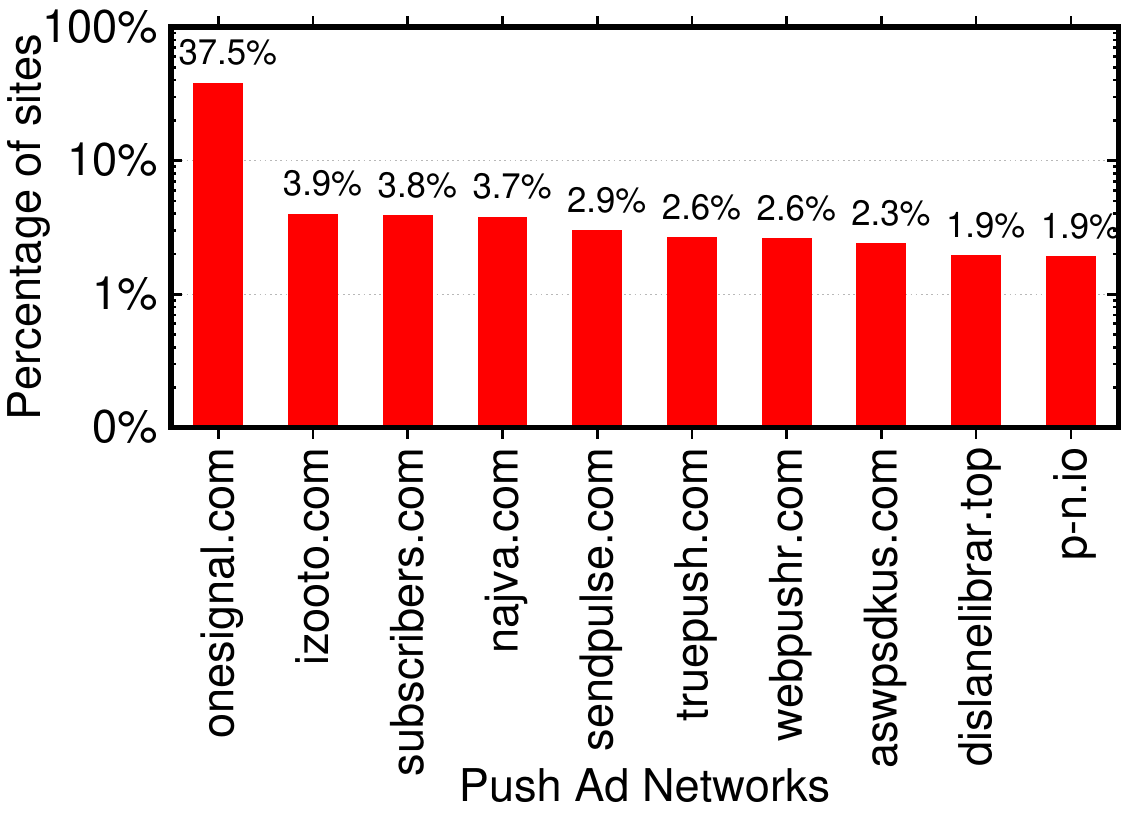}
        \caption{Portion of unique sites collaborating with each of the top ad servers in our dataset. \textit{onesignal.com} dominates the market (37.49\%) with the majority of the rest of the ad servers owning less than 4\% each (note: y-axis in log-scale).}
        \label{fig:topAdvertisers}
    \end{minipage}
    \hfill
\end{figure}

\point{What is the communications between \sws and Web?}
Next, by analyzing the traffic that the registered \sws generate, we see that \swFirstPartyperc communicate only with the first party.
However, \swThirdPartyperc of them communicate with at least one 3rd party, and \swNoPartyperc of them do not communicate with the web at all (Table~\ref{tab:summary}).
In Figure~\ref{fig:thirdParties}, we plot the number of distinct 3rd parties each registered \sw in our dataset communicates with.
As we can see, \noTp of the \sws communicate with no 3rd party (as mentioned: \swFirstPartyperc connects with the first party only, and \swNoPartyperc with no one), when the majority (\oneTp) communicates with exactly one 3rd party, proving that there are specific agreements between publishers and 3rd party advertisers, analytics, content or library providers.
It is important to note that \emph{there is a significant \twoMoreTp communicating with 2 or more (even 26!) distinct 3rd parties.}

\point{How much do \sws support Push Advertising?}
By using the popular filter-list of \emph{1Hosts}, we classify the type of domains the \sws connect with in our dataset.
Surprisingly, as we see in Figure~\ref{fig:NumOfAdServers}, in essence \emph{3rd party communications of \sws are used for advertising}, since the majority (\swtpToAdsperc) of the \sws that connect with 3rd parties, establish these connections to receive content from at least one push advertiser (9.51\% receive content from 2 advertisers or more).
On the contrary, only \swtpNOAdsperc of the \sws perform at least one request to 3rd parties, but communicate with zero ad servers.

\point{What is the popularity of sites leveraging Push Advertising?}
In Figure~\ref{fig:rank}, we plot the popularity rank of the websites that deploy \sws on the users' side.
As we see, the sites that tend to deploy ad pushing \sws are of lower popularity ranks in comparison to the ones that use \sw only locally, without connecting to any remote server.
Specifically, the median site that registers \sw that does not connect with any remote server has a popularity rank of around 40000 in Tranco (grey).
On the other hand, the median site with \sw that connects (i) only with first party domains, or with 3rd parties that do not include ads is around 50000 rank (green or orange) (ii) with push ad domains around 60000 rank (red).

Also, as we see in Figure~\ref{fig:sitesSWcontent} (in red) where we consider only sites with \sws that communicate with ad servers, the content categories are topped by `News and Media' (27\%), `Computers, Electronics and Technology'(6.42\%) and `Adult' (4.79\%).
This is somewhat expected, since `News and Media' sites have higher chances to convince a user to give their consent to receive timely news updates via push notifications, that can also include ads.

In Figure~\ref{fig:adcontent}, we further analyze these content categories by selecting their sites in our dataset that communicate with 3rd parties via their deployed \sws.
Then, we measure what portion of them does that for advertising purposes.
We see that \textbf{Soccer} sites lead this effort, with a percentage close to 85\%.
This means that from all Soccer sites using \sws to communicate with 3rd parties, 85\% use them to communicate specifically with at least one ad server.
The `Animation and Comics` follow closely with 84\% and `File Sharing and Hosting'
are next, with 83.78\%.
The `News and Media' are in fifth place with a bit more than 77\%.
These high portions suggest that 75-80\% of these websites use \sws for ads.

One can not help but wonder why were \sws invented in the first place.
It is true that several people may argue that \sws were invented to provide offline 
operation, synchronize data in the background, and retrieve updates.
However, we see a different picture here: \sws that communicate with 3rd parties are primarily used for advertisements, thus opening a new way to reach users' desktop: a way invented for a different purpose.
Even if these push notifications require user to give consent, the website is free to abuse this consent at any time by delivering ad messages instead of the news updates the user was interested in receiving. These ad messages appear via the \sws as native ads and cannot be controlled (or filtered out) by ad-blockers. 
One can only smile in melancholy at the Google Developers guide advising: ``\emph{Whatever you do, do not use notifications for advertising of any kind.}'' \footnote{https://developers.google.com/web/ilt/pwa/introduction-to-push-notifications}

\begin{figure}[t]
    \centering
    \includegraphics[width=.6\textwidth]{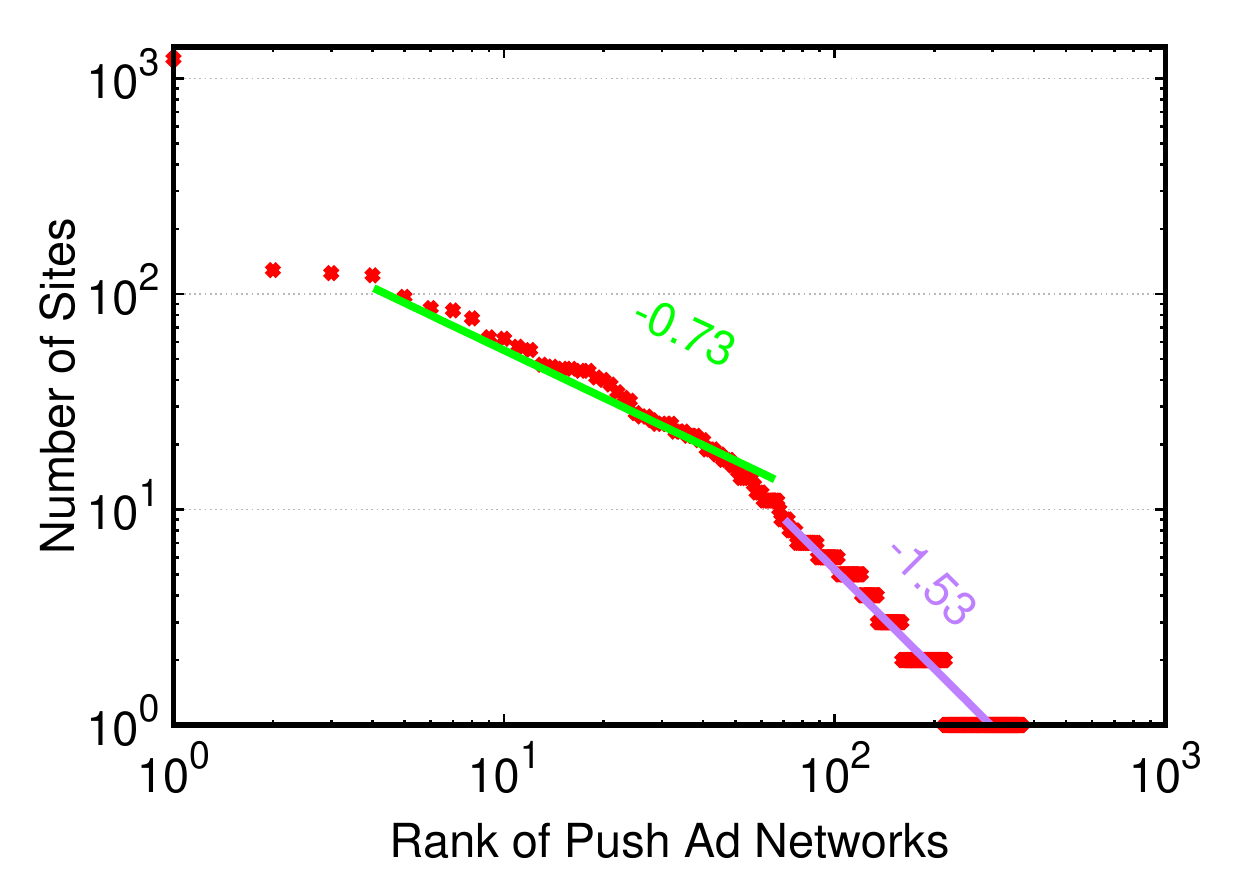}
    \caption{Distribution of the number of sites that each Push Advertiser in our dataset delivers ad notifications to. The points in the plot tend to converge to two straight lines for large numbers in the x axis, following a piece-wise power-law distribution.}
    \label{fig:advertisers}
\end{figure}

\point{Which are the dominant Push Ad Networks?}
In Figure~\ref{fig:topAdvertisers}, we plot the top 10 most popular Push Ad Networks in our data, and the portion of the registered \sws they communicate with.
We see that \textit{onesignal.com} dominates push advertising by owning more than 37.49\% of the market, with the majority of the rest Push Ad Networks owning less than 4\% each.
In Figure~\ref{fig:advertisers}, we plot the distribution of all push ad networks in our dataset along with the sites they deliver push ads to.
We can see that the distribution can be modeled by two straight lines for large numbers in the x-axis, indicating that the distribution has a piece-wise power-law tail.
We can also see the head representing the major player \textit{onesignal.com}.


\section{Related Work}
\label{sec:related}
The powerful technology of Service Workers provides rich functionality to developers and has triggered an important body of research around its security and privacy aspects.
Papadopoulos \etal in~\cite{papadopoulosmaster} are the first to study \sws in an attempt to raise awareness regarding a new class of attacks that exploit this exact HTML 5 functionality.
Specifically, the authors investigated the potential security vulnerabilities of \sws and they demonstrated multiple attack scenarios from cryptojacking to malicious computations (\eg distributed password cracking), as well as Distributed Denial of Service attacks.

Karami \etal in~\cite{karami2021awakening} studied attacks that aim to exploit \sws vulnerabilities to ex-filtrate important privacy information from the user. 
Specifically, they demonstrated two history-sniffing attacks that exploit the lack of appropriate isolation in these browsers including a non-destructive cache-based version.
Finally, the authors proposed a countermeasure and developed a tool that streamlines its deployment, thus facilitating adoption at a large scale.

Chinprutthiwong \etal in~\cite{chinprutthiwong2020security} described a novel Service Worker-based Cross-Site Scripting (SW-XSS) attack inside a \sw, that allows an attacker to obtain and leverage \sw privileges.
Additionally, they developed a SW Scanner to analyze top websites in the wild, and they found 40 websites vulnerable to this attack including several popular and high ranking websites.
Squarcina \etal in~\cite{squarcinaremote} demonstrated how a traditional XSS attack can abuse the Cache API of a \sw to escalate into a person-in-the-middle attack against cached content, thus, compromising its confidentiality and integrity.

Subramani \etal in~\cite{10.1145/3419394.3423631} proposed PushAdMiner: a new tool to detect Web Push Notifications (WPNs) on the Web.
Contrary to our work, the authors focus only on ad related WPNs messages by collecting and analyzing 21,541 WPN messages and 572 ad campaigns, for a total of 5,143 WPN-based ads reporting 51\% of them as malicious.
%
Finally, Lee \etal in~\cite{lee2018pride} conducted a systematic study of the security and privacy aspects of PWAs.
They demonstrated a cryptojacking and a browser history exfiltration attack.
They also suggested possible mitigation measures against the vulnerabilities of PWAs and corresponding \sws.
\section{Summary \& Conclusion}
\label{sec:conclusion}

In this paper, we set out to explore the ecosystem of \swslong and how websites overuse them to deliver ads (even when user has deployed ad-blockers).
We analyzed the top \topList websites of the Tranco list and our findings can be summarized as follows:

\begin{enumerate}[itemsep=0pt,topsep=0pt]
    \item A non-trivial percentage (\sitesWithSWperc) of sites deploy a \sw on the user side.
    \item Within a period of 5 months (12.20-05.21), there has been a \adoptionIncrease increase in the adoption of \sws.
    \item Overall, by using Wayback Machine, we found that from 2015 till today, there were $1.62\times$ more publishers per year, on average, utilizing \sws in their web applications.
    \item \noTp of the \sws communicate with no 3rd party (\swFirstPartyperc connects with its first party only and \swNoPartyperc connects with nobody).
    The majority (\oneTp) communicates with exactly one 3rd party with a significant \twoMoreTp communicating with 2 or more (and up to 26) distinct 3rd parties.
   \item  Third-party communications are mostly for pushing ads: {\bf A stunning \swtpToAdsperc of the registered \sws that communicates with 3rd party servers, communicate with at least one advertiser}.
   \item Most of the ads-pushing \sws are deployed on `News and Media' related sites (27\%), with the `Computers, Electronics and Technology' (6.42\%), and `Adult' (4.79\%) related sites following.
   \item For some website categories such as `Soccer' and `File Sharing', the percentage of ads-pushing \sws reaches as high as 85\%. 
\end{enumerate}

Our study on \swslong has revealed several surprising results with respect to the use of \sws on Web applications and websites.
Future research could look into leakage of user personal information and tracking from \sws, as well as how ad-blockers can be revamped to still provide effective ad-filtering to their end-users.

\section*{Acknowledgements}
This project received funding from the EU H2020 Research and Innovation programme under grant agreements No 830927 (Concordia), No 830929 (CyberSec4Europe), No 871370 (Pimcity) and No 871793 (Accordion).
These  results reflect only the authors' view and the Commission is not responsible for any use that may be made of the information it contains.

\bibliographystyle{unsrt}
\bibliography{main}
\end{document}